\documentclass[%
 reprint,
 amsmath,amssymb,
 aps,
 pra,
 floatfix,
]{revtex4-2}

\usepackage{graphicx}
\usepackage{dcolumn}
\usepackage{bm}
\usepackage{amsmath}   
\usepackage{braket}    
\usepackage{booktabs}  
\usepackage{cancel}    
\usepackage{color}     
\usepackage{hyperref}  
\usepackage{lineno}    
\usepackage{lipsum}    
\usepackage{longtable} 
\usepackage{multirow}  
\usepackage{pifont}    
\usepackage{relsize}   
\usepackage{stmaryrd}  
\usepackage{xcolor}    
\usepackage[T1]{fontenc}
\usepackage{soul}

\newcommand{\kolora}{black}
\begin{document}

\title{Spectral line shape in the limit of frequent velocity-changing collisions}

\author{Nikodem Stolarczyk} 
\email{NikodemStolarczyk319@gmail.com}
\author{Piotr Wcisło}
\author{Roman Ciuryło}
\affiliation{Institute of Physics, Faculty of Physics, Astronomy and Informatics, Nicolaus Copernicus University in Toru\'n, ul Grudzi\k{a}dzka 5, 87-100 Toruń, Poland}

\date{\today}

\begin{abstract}
The speed-dependent spectral line profiles collapse into a simple Lorentz profile in the regime dominated by the velocity-changing collisions. We derive general formulas for the effective width and shift of the Lorentzian for arbitrary speed-dependent collisional broadening and shift and velocity-changing collision operators. For a quadratic speed dependence of collisional broadening and shift, and the billiard ball model of velocity-changing collisions, we provide simple analytical expressions for the effective Lorentzian width and shift. We show that the effective Lorentzian width and shift split into components originating from the well-known Dicke-narrowed Doppler width, speed-averaged collisional broadening and shift, their speed dependencies, and a product term that mixes the contributions of the broadening and shift speed dependencies. We show how the components depend on rates of speed-changing and velocity-changing collisions related to the perturber/absorber mass ratio. We validate analytical formulas numerically on the example of H$_{2}$ transition perturbed by He.

\end{abstract}

\maketitle

\section{Introduction}
\label{sec:intro}

The shape of molecular spectral line affected by Doppler broadening and absorber-perturber collisions, in the general case, requires numerical evaluation and cannot be represented by a simple analytical function, unless some simplifications or assumptions are made~\cite{Hartmann2008book}. The goal of this work is to show that, in the case when the velocity-changing collisions dominate other line-shape effects, the spectral line shape collapses to an ordinary Lorentz profile and the expressions for its width and shift can be provided analytically.  
\par
The Lorentz profile has been used for over a century to describe collisionally-broadened atomic and molecular spectral lines. It is particularly justified at high pressure of perturbers, much lighter than absorbers, in microwave spectral range, where the speed dependence of collisional broadening and shift~\cite{Berman1972,Ward1974}, as well as Doppler broadening, can be neglected. Interestingly, also the Gaussian shape of the spectral line can collapse to the Lorentz profile when the velocity-changing collisions reduce the mean free path of the absorber well below the wavelength of absorbed radiation as predicted by Dicke~\cite{Dicke1953,Wittke1956} in the fifties. It was later demonstrated numerically~\cite{Ciurylo2002BB}, that the weighted sum of Lorentz profiles (WSL)~\cite{Pickett1980,Farrow1989}, under frequent velocity-changing collisions described by the billiard-ball model~\cite{Lindenfeld1979,Lindenfeld1980} approaches the Lorentz profile and the convergence is faster for lower perturber/absorber mass ratio. It is related to relative contribution of speed change during velocity-changing collisions which is determined by perturber/absorber mass ratio~\cite{Hoang2002,Ciurylo2002H2}.  
Recently, studies on line shapes for which width is dominated by the speed-dependent collisional shift led to formulating a simple analytical expression~\cite{Wcislo2016,Martinez2018} (see Eq.~(15) in Ref.~\cite{Wcislo2016}) for width of Lorentzian profile approximating such line shape when speed-dependent collisional shift is described by a quadratic function~\cite{Rohart1994} and velocity-changing collisions are approximated by the hard-collision model~\cite{Nelkin1964,Rautian1967}. The phenomenological finding from Ref.~\cite{Wcislo2016} got justification in a derivation~\cite{Stolarczyk2022} coming from the speed-dependent hard-collision profile~\cite{Lance1997,Pine1999}.
\par
In this article, we generalize the results from Ref.~\cite{Stolarczyk2022}; we derive general formulas for the effective width and shift of a Lorentzian to which a sophisticated line shape model, based on any arbitrary speed-dependent collisional broadening and shift and velocity-changing collision operator, converges in the limit of frequent velocity-changing collisions. For quadratic speed dependencies of collisional width and shift, and the billiard ball model of velocity-changing collisions, we provide a simple analytical expression for the effective Lorentzian width and shift. We show how their components depend on rates of speed-changing and velocity-changing collisions related to the perturber/absorber mass ratio. We validate the analytical formulas numerically.

\section{Algebraic representation of a spectral line shape}
\label{sec:algebraic}

In general, the shape of an isolated spectral line affected by Doppler broadening and collisions with perturbers can be evaluated~\cite{Blackmore1987,Shapiro2002} from a function $h(\omega,\vec{v})$, 
\begin{equation}
\label{eq:defI}
I(\omega)=\frac{1}{\pi} \mbox{\rm Re}\;\left\{(1,h(\omega,\vec{v}))\right\} ,
\end{equation}
where $(\cdot,\cdot)$ is defined as a product $(a(\vec{v}),b(\vec{v}))=\int d^{3}\vec{v} f_{m_A}(\vec{v}) a(\vec{v}) b(\vec{v})$ of two functions $a(\vec{v})$, $b(\vec{v})$ of absorber velocity $\vec{v}$, $f_{m_{A}}(\vec{v})=(\pi v^{2}_{m_{A}})^{-3/2} \exp(-v^{2}/v^{2}_{m_{A}})$ is Maxwellian distribution, $v_{m_{A}}=\sqrt{k_{B}T/(2m_{A})}$ is the most probable speed of the absorber having mass $m_{A}$ at temperature $T$, and $k_{B}$ is Boltzmann’s constant. Function $h(\omega,\vec{v})$ fulfils transport-relaxation kinetic equation~\cite{Shapiro2002},
\begin{equation}
\label{eq:KinEq}
1=-i(\omega-\omega_{0}-\vec{k}\cdot\vec{v})h(\omega,\vec{v})-\hat{S}^{f}h(\omega,\vec{v}),
\end{equation}
where $\omega_{0}$ is unperturbed transition frequency, $\vec{k}$ is a wavevector of radiation, and operator $\hat{S}^{f}$ describes the effect of collisions with perturbers. 
\par
Equations (\ref{eq:defI}) and (\ref{eq:KinEq}) can be converted into an algebraic form expanding function $h(\omega,\vec{v})$,
\begin{equation}
\label{eq:SumPhi}
h(\omega,\vec{v})=\sum_{s=0}^{\infty}c_{s}(\omega)\varphi_{s}(\vec{v}),
\end{equation}
in a set of orthonormal functions fulfilling condition $(\varphi_{s}(\vec{v}),\varphi_{s'}(\vec{v}))=\delta_{s,s'}$. We set $\varphi_{0}(\vec{v})=1$. The expansion coefficients $c_{s}(\omega)$ depend only on frequency for given operator $\hat{S}^{f}$. In this basis, any operator $\hat{A}$ can be represented by matrix $\mbox{\bf A}$, having matrix elements
$[\mbox{\bf A}]_{s,s'}=(\varphi_{s}(\vec{v}),\hat{A}\varphi_{s'}(\vec{v}))$.
\par
The shape of an isolated spectral line can have algebraic representation by a series of Lorentz profiles. Following the approaches from~\cite{Blackmore1987,Lindenfeld1980,Podivilov1994,Robert1998} and notation described
in Refs.~\cite{Shapiro2002,Ciurylo2002icsls,Ciurylo2002BB}, the line shape can be written as~\cite{Wehr2002,Ciurylo2002icsls,Ciurylo2002BB}:
\begin{equation}
\label{eq:defIc0}
I(\omega)=\frac{1}{\pi} \mbox{\rm Re}\;\left\{c_{0}(\omega)\right\} ,
\end{equation}
where  the coefficient
$c_{0}(\omega)$ can be evaluated by solving a set of complex linear equations,
\begin{equation}
\label{eq:EqL}
\mbox{\bf \underline{b}} = \mbox{\bf L}(\omega) \mbox{\bf \underline{c}}(\omega) ,
\end{equation}
for the coefficients $c_{s}(\omega)$, where $s=0, 1, ..., s_{\rm max}$.
Here the column $\mbox{\bf \underline{b}}$
contains unity in the position 0 and zeros in other positions,
i.e. $[\mbox{\bf \underline{b}}]_{s}=\delta_{0,s}$,
and the column $\mbox{\bf \underline{c}}(\omega)$ consists of
the coefficients $c_{s}(\omega)$,
i.e. $[\mbox{\bf \underline{c}}(\omega)]_{s}=c_{s}(\omega)$.
The matrix $\mbox{\bf L}(\omega)$ depends on the frequency $\omega$
and has the following form:
\begin{equation}
\label{eq:defL}
\mbox{\bf L}(\omega)=-i(\omega-\omega_{0})\mbox{\bf 1}+i\mbox{\bf K}- \mbox{\bf S}^{f},
\end{equation}
where $\omega_{0}$ corresponds to the unperturbed frequency of the transition,
$\mbox{\bf 1}$ is the unit matrix,
$[\mbox{\bf 1}]_{s,s'}=\delta_{s,s'}$,
$\mbox{\bf K}$ is the matrix that represents the Doppler shift,
$\mbox{\bf S}^{f}=\mbox{\bf S}_{D}^{f}+\mbox{\bf S}_{VC}^{f}$ is the matrix
which represents the collision operator split into two components:
$\mbox{\bf S}_{D}^{f}$ is the matrix which represents
the "dephasing and relaxation"~\cite{Wcislo2018,Slowinski2020} collisional width and shift and $\mbox{\bf S}_{VC}^{f}$ is the matrix which represents the "velocity-changing" collision operator also affected by dephasing and relaxation. The representation used here has a property $[\mbox{\bf S}_{VC}^{f}]_{s,0}=[\mbox{\bf S}_{VC}^{f}]_{0,s}=0$ for any $s$ and $[\mbox{\bf K}]_{0,0}=0$. Moreover, all matrices discussed in this work are symmetric.
\par
In practice, the coefficient $c_{0}(\omega)$ is calculated using the
diagonalization technique (c.f.~\cite{Robert1998,Dolbeau1999,Shapiro2001Dicke}). To do it in this way, one needs to find the full set of eigenvectors, $\mbox{\bf \underline{e}}_{j}$, and corresponding eigenvalues $\varepsilon_{j}$ which fulfill the following equation:
\begin{equation}
\label{eq:EqEigen} (i\mbox{\bf K}- \mbox{\bf S}^{f}) \mbox{\bf
\underline{e}}_{j}=\varepsilon_{j} \mbox{\bf \underline{e}}_{j} ,
\end{equation}
where $j=0, 1, ..., s_{\rm max}$.
Once the eigenvectors and eigenvalues are known, the coefficient $c_{0}(\omega)$ can be computed from the following expression:
\begin{equation}
\label{eq:SumLor}
c_{0}(\omega)=\sum_{j=0}^{s_{\rm max}}\frac{\beta_{j}[\mbox{\bf
\underline{e}}_{j}]_{0}}{\varepsilon_{j}-i(\omega-\omega_{0})}  ,
\end{equation}
where the coefficients $\beta_{j}$ fulfill the relation $\mbox{\bf \underline{b}}=\sum_{j=0}^{s_{\rm max}}\beta_{j}\mbox{\bf
\underline{e}}_{j}$. The main advantage of this approach is that the time-consuming diagonalization can be carried out once and this is sufficient to calculate the whole line shape.

\section{High frequency of the velocity-changing collisions limit}
\label{sec:derivation}
Our derivation is carried out in the limit where the velocity-changing collisions dominate over the Doppler broadening and collisional broadening and shift. We assume the absolute values of all matrix elements of $\mbox{\bf K}$ and $\mbox{\bf S}_{D}^{f}$, as well as detuning, $\omega-\omega_{0}$, to be much smaller than the absolute value of the effective optical frequency of velocity changing collisions, $\nu_{\rm opt}$. Importantly, $\nu_{\rm opt}$ can be complex due to the dephasing associated with optical velocity-changing collisions. All nonzero matrix elements of $\mbox{\bf S}_{VC}^{f}$ are directly proportional to $\nu_{\rm opt}$. 
{\color{\kolora}
The recognition of $\nu_{\rm opt}$ as a complex quantity was originally put forth by Rautian and Sobelmann \cite{Rautian1967}. Subsequent support for this notion came from the comparison between measurements~\cite{Pine1994} and theoretical estimations~\cite{Demeio1995}. The complex form of $\nu_{\rm opt}$ has been justified through both semiclassical approaches~\cite{Rautian1967, Nienhuis1978, Looney1987, Ciurylo2001JQSRT} and quantum treatments~\cite{Hess1972, Kochanov1977, Nienhuis1978, Berman1982, Berman1984, Monchick1986, Thibault2017}, see also the references cited therein.
}
\par
For the matrix $\mbox{\bf S}_{VC}^{f}$ we can calculate eigenvectors, $\mbox{\bf \underline{e}}_{j}^{VC}$, and eigenvalues, $\varepsilon_{j}^{VC}$, which fulfil the following equation:
\begin{equation}
\label{eq:EqEigenVC} 
- \mbox{\bf S}_{VC}^{f} \mbox{\bf
\underline{e}}_{j}^{VC}=\varepsilon_{j}^{VC} \mbox{\bf \underline{e}}_{j}^{VC} .
\end{equation}
It is easy to see that one of the eigenvectors of the matrix having property $[\mbox{\bf S}_{VC}^{f}]_{s,0}=[\mbox{\bf S}_{VC}^{f}]_{0,s}=0$ is vector $\mbox{\bf \underline{b}}$ and its corresponding eigenvalue is zero. Therefore, we can set $\mbox{\bf \underline{e}}_{0}^{VC}=\mbox{\bf \underline{b}}$ and $\varepsilon_{0}^{VC}=0$. Consequently, for $j\neq 0$ we have $[\mbox{\bf \underline{e}}_{j}^{VC}]_{0}=0$ and $\varepsilon_{j}^{VC}\sim \nu_{\rm opt}$.
\par
Now we can rewrite Eq.~(\ref{eq:EqEigen}) in the form:
\begin{equation}
\label{eq:EqEigenVCDK} (-\mbox{\bf S}^{f}_{VC}-\mbox{\bf S}^{f}_{D}+i\mbox{\bf K}) \mbox{\bf
\underline{e}}_{j}=\varepsilon_{j} \mbox{\bf \underline{e}}_{j} ,
\end{equation}
where the matrix, which we want to diagonalize, is split into the dominating part $\mbox{\bf S}^{f}_{VC}$ and the perturbation part containing $\mbox{\bf S}^{f}_{D}$ and $\mbox{\bf K}$. We take advantage of that and approximate eigenvectors $\mbox{\bf \underline{e}}_{j}$ by $\mbox{\bf \underline{e}}_{j}^{VC}$ and eigenvalues $\varepsilon_{j}$ by $\varepsilon_{j}^{VC}$ and improve them by perturbation corrections. Setting $\mbox{\bf \underline{e}}_{j}\approx\mbox{\bf \underline{e}}_{j}^{VC}$ we get $\beta_{j}=\delta_{j,0}$. In this, way Eq.~(\ref{eq:SumLor}) can be approximated with the Lorentz profile:
\begin{equation}
\label{eq:cLor}
c_{0}(\omega)\approx\frac{1}{\varepsilon_{0}-i(\omega-\omega_{0})},
\end{equation}
where the eigenvalue $\varepsilon_{0}$ is approximated by second-order perturbation. The eigenvalues, $\varepsilon_{j}$, in the second-order perturbation are given by the following expression:
\begin{multline}
\label{eq:Per2nd}
\varepsilon_{j}\approx\varepsilon_{j}^{VC}
+\sum_{s,s'=0}^{s_{\rm max}} [\mbox{\bf \underline{e}}_{j}^{VC}]_{s}
(-[\mbox{\bf S}^{f}_{D}]_{s,s'}+i[\mbox{\bf K}]_{s,s'})
[\mbox{\bf \underline{e}}_{j}^{VC}]_{s'}+\\
+\sum_{\substack{j'=0\\j'\neq j}}^{s_{\rm max}}
\frac{(
\sum_{s,s'=0}^{s_{\rm max}}
[\mbox{\bf \underline{e}}_{j}^{VC}]_{s}
(-[\mbox{\bf S}^{f}_{D}]_{s,s'}+i[\mbox{\bf K}]_{s,s'})
[\mbox{\bf \underline{e}}_{j'}^{VC}]_{s'}
)^{2}}
{\varepsilon_{j}^{VC}-\varepsilon_{j'}^{VC}}.
\end{multline}
This equation for $\varepsilon_{0}$ simplifies to the following form
\begin{multline}
\label{eq:Eps0}
\varepsilon_{0}\approx
-[\mbox{\bf S}^{f}_{D}]_{0,0}+\\
+\sum_{j'=1}^{s_{\rm max}}
\frac{\{
\sum_{s'=1}^{s_{\rm max}}
(-[\mbox{\bf S}^{f}_{D}]_{0,s'}+i[\mbox{\bf K}]_{0,s'})
[\mbox{\bf \underline{e}}_{j'}^{VC}]_{s'}
\}^{2}}
{-\varepsilon_{j'}^{VC}},
\end{multline}
remembering that $\varepsilon_{0}^{VC}=0$, $[\mbox{\bf \underline{e}}_{0}^{VC}]_{s}=\delta_{0,s}$ and $[\mbox{\bf K}]_{0,0}=0$.
In this way, we get a single Lorentz profile,
\begin{equation}
\label{eq:defI0}
I(\omega)=\frac{1}{\pi}
\frac{\Gamma_{\rm eff}}
{\Gamma_{\rm eff}^{2}+(\omega-\omega_{0}-\Delta_{\rm eff})^{2}},
\end{equation}
in the limit of frequent velocity-changing collisions, here
\begin{equation}
\label{eq:defGD}
\Gamma_{\rm eff}+i\Delta_{\rm eff}=\varepsilon_{0}.
\end{equation}

It should be noted that we can always use the basis in which matrix $\mbox{\bf S}^{f}_{VC}$ is diagonal. In such case $[\mbox{\bf \underline{e}}_{j}^{VC}]_{s}=\delta_{j,s}$,
$\varepsilon_{j}^{VC}=-[\mbox{\bf S}^{f}_{VC}]_{j,j}$ and Eq.~(\ref{eq:Eps0}) takes simple form
\begin{equation}
\label{eq:Eps0diag}
\varepsilon_{0}\approx
-[\mbox{\bf S}^{f}_{D}]_{0,0}
+\sum_{s=1}^{s_{\rm max}}
\frac{
(-[\mbox{\bf S}^{f}_{D}]_{0,s}+i[\mbox{\bf K}]_{0,s})^{2}
}
{[\mbox{\bf S}^{f}_{VC}]_{s,s}}.
\end{equation}
Furthermore, one can use a basis in which if a matrix element is nonzero, $[\mbox{\bf S}^{f}_{D}]_{0,s}\neq 0$, then the corresponding matrix element is zero, $[\mbox{\bf K}]_{0,s}=0$, and, if $[\mbox{\bf K}]_{0,s}\neq 0$ then $[\mbox{\bf S}^{f}_{D}]_{0,s}=0$. It is a simple consequence of the symmetry of the corresponding operators. The collisional broadening and shift depend on the absolute value of the absorber velocity, $v=|\vec{v}|$, and are not dependent on velocity direction. On the other hand, the Doppler shift is proportional to the scalar product $\vec{k}\cdot\vec{v}$ which depends on velocity direction. Taking this into account, we can rewrite Eq.~(\ref{eq:Eps0diag}) in the following form:
\begin{equation}
\label{eq:Eps0diag2}
\varepsilon_{0}\approx
-[\mbox{\bf S}^{f}_{D}]_{0,0}
-\sum_{s=1}^{s_{\rm max}}
\frac{
[\mbox{\bf K}]_{0,s}^{2}
}
{[\mbox{\bf S}^{f}_{VC}]_{s,s}}
+\sum_{s=1}^{s_{\rm max}}
\frac{
[\mbox{\bf S}^{f}_{D}]_{0,s}^{2}
}
{[\mbox{\bf S}^{f}_{VC}]_{s,s}}.
\end{equation}

\section{Application in case of quadratic speed-dependent collisional broadening and shift}
\label{sec:application}

To get physical insight into the expressions derived above, we consider matrix representation using Burnett functions~\cite{Lindenfeld1979} described in Appendix~\ref{appendix:a}. We assume quadratic speed dependence of collisional broadening and shift~\cite{Rohart1994}. It means that the operator
\begin{equation}
\label{eq:DefSD}
\hat{S}_{D}^{f}=-\Gamma(v)-i\Delta(v)
\end{equation}
is determined by $\Gamma(v)$ and $\Delta(v)$ given in the following form
\begin{multline}
\label{eq:DefGD2}
\Gamma(v)+i\Delta(v)= \Gamma_{0}+i\Delta_{0} +(\Gamma_{2}+i\Delta_{2})\left(\frac{v^{2}}{v_{m}^{2}}-\frac{3}{2}\right),
\end{multline}
where $\Gamma_{0}$ and $\Delta_{0}$ are the collisional broadening and shift parameters, averaged over absorber velocity, respectively. Quadratic speed dependencies of collisional broadening and shift are described by $\Gamma_{2}$ and $\Delta_{2}$ parameters, respectively. 
\par
Discussing the velocity-changing collisions, we will focus on the case where matrix $\mbox{\bf S}^{f}_{VC}$ is diagonal or is approximated by a diagonal matrix.
\par
In the Burnett functions basis representation, instead of the index $s=0, 1, ..., s_{\rm max}$, we prefer to use two other indices, $n=0, 1, ..., n_{\rm max}$ and $l=0, 1, ..., l_{\rm max}$. The pair of indices $nl$ can be connected with $s=n+(n_{\rm max}+1)l$ and $s_{\rm max}=n_{\rm max}+(n_{\rm max}+1)l_{\rm max}$.

The properties of matrix elements in Burnett functions representation are summarized in Appendix~\ref{appendix:a}. The assumptions made above constrain the number of nonzero matrix elements which contribute to Eq.~(\ref{eq:Eps0diag2}) to only a few,
\begin{equation}
\label{eq:Eps02}
\varepsilon_{0}=
-[\mbox{\bf S}^{f}_{D}]_{00,00}
-
\frac{[\mbox{\bf K}]_{00,01}^{2}}
{[\mbox{\bf S}^{f}_{VC}]_{01,01}}
+
\frac{[\mbox{\bf S}^{f}_{D}]_{00,10}^{2}}
{[\mbox{\bf S}^{f}_{VC}]_{10,10}}.
\end{equation}
Now we can explicitly express the matrix elements: $[\mbox{\bf K}]_{00,01}=\omega_{D}\sqrt{1/2}$, where $\omega_{D}=kv_{m_{A}}$, $[\mbox{\bf S}^{f}_{D}]_{00,00}=-(\Gamma_{0}+i\Delta_{0})$,
and $[\mbox{\bf S}^{f}_{D}]_{00,10}=-(\Gamma_{2}+i\Delta_{2})\sqrt{3/2}$. 

\subsection{Hard collision model}
\label{sec:HC}

To discuss the velocity-changing collisions we first consider the hard collision (HC) model~\cite{Nelkin1964,Rautian1967}, discussed recently in Ref.~\cite{Stolarczyk2022} in a similar context. The HC model is frequently used due to its simplicity, despite its shortcomings. In many situations, it helps to get some analytical results. The operator $\hat{S}_{HC}^{f}$ describing hard velocity changing collisions has the following form:
\begin{equation}
\label{eq:defHC}
\hat{S}_{HC}^{f}h(\omega,\vec{v})=
-\nu_{\rm opt}h(\omega,\vec{v})
+\nu_{\rm opt}\int d^{3}\vec{v}'f_{m_{A}}(\vec{v}')
h(\omega,\vec{v}'),
\end{equation}
where $\nu_{\rm opt}$ is the effective optical frequency of velocity-changing collisions, which, in the general case, can be a complex number~\cite{Rautian1967,Hess1972,Pine1994,Demeio1995}.
It can be shown that in any orthonormal base assuming $\varphi_{0}(\vec{v})=1$, the matrix representation of the $\hat{S}_{HC}^{f}$ operator is diagonal, $[\mbox{\bf S}^{f}_{HC}]_{s,s'}=-\nu_{\rm opt}\delta_{s,s'}(1-\delta_{0,s})$. All its diagonal elements are equal to $-\nu_{\rm opt}$, except $[\mbox{\bf S}^{f}_{HC}]_{0,0}=0$. Therefore, also in the Burnett functions representation, we can write that $[\mbox{\bf S}^{f}_{HC}]_{01,01}=[\mbox{\bf S}^{f}_{HC}]_{10,10}=-\nu_{\rm opt}$. Inserting these matrix elements into Eqs.~(\ref{eq:Eps02}) and~(\ref{eq:defGD}) we got 
\begin{equation}
\label{eq:GDHC}
\Gamma_{\rm eff}^{HC}+i\Delta_{\rm eff}^{HC}=
\Gamma_{0}+i\Delta_{0}
+\frac{\omega_{D}^{2}}{2\nu_{\rm opt}}
-\frac{3\Gamma_{2}^{2}}{2\nu_{\rm opt}}
+\frac{3\Delta_{2}^{2}}{2\nu_{\rm opt}}
-i\frac{3\Gamma_{2}\Delta_{2}}{\nu_{\rm opt}}.
\end{equation}
The above result has also been obtained in our recent paper with a different method, 
see Eqs.~(19a)--(19d) in Ref.~\cite{Stolarczyk2022}, where we have shown that speed-dependent hard-collision profile~\cite{Pine1999} collapses into a simple Lorentz profile in the limit of frequent velocity-changing collisions.

\subsection{Soft collision model}
\label{sec:SC}
In the case when perturbers are much lighter then absorbers, the velocity changing collisions are described by soft collision (SC) model \cite{Chandrasekhar1943}. This model was introduced by Galatry \cite{Galatry1961} into the theory of Dicke narrowed spectral line shapes. By that time, the Galatry profile (GP) became one of the most frequently used expression to describe collisionally-narrowed spectra. The exact speed-dependent Galatry profile (SDGP) with quadratic speed dependence of collisional width and shift was given in Ref.~\cite{Ciurylo2001SDGP} and should be not confused with its approximated expression provided by Prime~\textit{et al}~\cite{Priem2000CO} using appproach from Ref. \cite{Ciurylo1997}. It was shown in Refs.~\cite{Lindenfeld1979,Lindenfeld1980} that the velocity changing soft collisions operator $\hat{S}_{HC}^{f}$ is represented with a diagonal matrix operator in the Burnett functions basis. The relevant matrix elements are given by the following expressions:  $[\mbox{\bf S}^{f}_{SC}]_{00,00}=0$, $[\mbox{\bf S}^{f}_{SC}]_{01,01}=-\nu_{\rm opt}$, $[\mbox{\bf S}^{f}_{SC}]_{10,10}=-2\nu_{\rm opt}$, see Refs.~\cite{Lindenfeld1979,Lindenfeld1980}. In contrast to the HC model, the matrix elements $[\mbox{\bf S}^{f}_{SC}]_{01,01}$ and $[\mbox{\bf S}^{f}_{SC}]_{10,10}$ are not identical. Inserting these matrix elements into Eqs.~(\ref{eq:Eps02}) and~(\ref{eq:defGD}) yields:
\begin{multline}
\label{eq:GDSC}
\Gamma_{\rm eff}^{SC}+i\Delta_{\rm eff}^{SC}=\\
\Gamma_{0}+i\Delta_{0}
+\frac{\omega_{D}^{2}}{2\nu_{\rm opt}}
-\frac{3\Gamma_{2}^{2}}{4\nu_{\rm opt}}
+\frac{3\Delta_{2}^{2}}{4\nu_{\rm opt}}
-i\frac{3\Gamma_{2}\Delta_{2}}{2\nu_{\rm opt}}.
\end{multline}
The three last terms of this equation are two times smaller then the ones derived in case of the HC model, Eq.~(\ref{eq:GDHC})~\cite{Stolarczyk2022}.
\begin{table*}[t]
    \centering
    \caption{Coefficients $f_D$, $f_{v^2}$ and their ratio $f_D/f_{v^2}$ evaluated for different $\alpha$ within billiard ball model.}
    \begin{tabular}{l l l l}
         $\alpha$& $f_D$& $f_{v^2}$& $f_D/f_{v^2}$  \\
         \hline
         0&1&1&1\\
         1/100&1.00000331561954&1.00000991995975&0.99999339572530\\
         1/50&1.00001319212471&1.00003935941918&0.99997383373542\\
         1/20&1.00008114950708 &1.00023998438455&0.99984120323127\\
         1/10&1.00031613831745 &1.00091992503335&0.99939676821212\\
         1/5&1.00119948726834&1.00336387252056&0.99784287105456\\
         1/4& 1.00182496341562&1.00501525070746&0.99682563295473\\
         1/3& 1.00310178272716&1.00822551613664&0.99491806810334\\
         1/2&1.00636139570140&1.01565719613513&0.99084750202223\\
         2/3&1.01027159637697&1.02349086931942&0.98708413202431\\
         1&1.01895378488101&1.03806540223391&0.98158919725889\\
         3/2& 1.03178515946401&1.05522030753985&0.97779122718887\\
         2&1.04294734723661&1.06757150938264&0.97693441429486\\
         3& 1.05994565861&1.0835515889&0.9782142996\\
         4&1.0717514674&1.093265697&0.980321134\\
         5&1.08028126&1.09975074&0.982296461\\
         10&1.101660&1.114423&0.9885474\\
         20&1.11523&1.122707&0.99334\\
         50&1.12464&1.1280&0.997\\
         100&1.1280&1.1280&1\\
         $\infty$&$32/9\pi\approx 1.131$&1.131&1\\
    \end{tabular}
    \label{tab:fd}
\end{table*}
\subsection{Billiard ball model}
\label{sec:BB}

The billiard ball (BB) model~\cite{Lindenfeld1980} provides a more realistic description of the velocity-changing collisions. This model properly accounts for the perturber/absorber mass ratio, $\alpha=m_{p}/m_{a}$. The matrix describing such velocity-changing collisions $[\mbox{\bf S}^{f}_{BB}]_{nl,n'l'}=\nu_{\rm opt}f_{D}M_{nl,n'l'}^{E*}$ is determined by coefficients $M_{nl,n'l'}^{E*}$ \cite{Lindenfeld1979,Lindenfeld1980} defined in Appendix~\ref{appendix:a}. This matrix is diagonal and the factor $f_{D}=1$ in case of $\alpha=0$, which corresponds to the soft collision (SC) model. For nonzero $\alpha$, the matrix $[\mbox{\bf S}^{f}_{BB}]_{nl,n'l'}$ is not diagonal and $f_{D}$ becomes greater than one, reaching $32/(9\pi)\approx 1.132$ for $\alpha=\infty$~\cite{Lindenfeld1980}. Nevertheless, we can approximate the original matrix by its diagonal simplification with the same diagonal elements $[\mbox{\bf S}^{f}_{BB}]_{nl,nl}$. The coefficients $M_{01,01}^{E*}=-1$ and $M_{10,10}^{E*}=-2M_{1}$, where $M_{1}=m_{A}/(m_{A}+m_{B})$ are of our particular interest since these allow to provide the explicit form of the matrix elements $[\mbox{\bf S}^{f}_{BB}]_{01,01}=-\nu_{\rm opt}f_{D}$ and $[\mbox{\bf S}^{f}_{BB}]_{10,10}=-\nu_{\rm opt}f_{D}2/(1+\alpha)$. In contrast to the HC model, these matrix elements are not identical. The matrix element $[\mbox{\bf S}^{f}_{BB}]_{01,01}\approx\nu_{\vec{v}}$ describes relaxation rate, $\nu_{\vec{v}}$, of the velocity vector, $\vec{v}$. On the other hand, $[\mbox{\bf S}^{f}_{BB}]_{10,10}\approx\nu_{v^{2}}$ describes relaxation rate, $\nu_{v^{2}}$, of $v^{2}$ which is directly related to speed, $v$. As it was shown in Refs.~\cite{Ciurylo2002H2,Wcislo2014}, the ratio between the speed, $v$- or its square, $v^{2}$-changing collisions and velocity, $\vec{v}$-changing collisions rate $\nu_{v^{2}}/\nu_{\vec{v}}=[\mbox{\bf S}^{f}_{BB}]_{10,10}/[\mbox{\bf S}^{f}_{BB}]_{01,01}=2/(1+\alpha)$ varies with $\alpha$. This ratio agrees well with results obtained from the classical molecular dynamics simulations based on realistic molecular interaction potentials~\cite{Hoang2002,Wcislo2014}. Inserting the matrix elements $[\mbox{\bf S}^{f}_{BB}]_{01,01}$ and $[\mbox{\bf S}^{f}_{BB}]_{10,10}$ into Eqs.~(\ref{eq:Eps02}) and~(\ref{eq:defGD}) we got
\begin{equation}
\label{eq:GDBB1}
\Gamma_{\rm eff}^{BB}+i\Delta_{\rm eff}^{BB}=
\Gamma_{0}+i\Delta_{0}
+\frac{\omega_{D}^{2}}{2\nu_{\vec{v}}}
-\frac{3\Gamma_{2}^{2}}{2\nu_{v^{2}}}
+\frac{3\Delta_{2}^{2}}{2\nu_{v^{2}}}
-i\frac{3\Gamma_{2}\Delta_{2}}{\nu_{v^{2}}}.
\end{equation}
\par
We do not need to limit our discussion to the diagonal approximation in this place. We can take into account the full matrix, $[\mbox{\bf S}^{f}_{BB}]_{nl,nl}$, and use Eq.~(\ref{eq:Eps0}). However, to derive the exact analytical expressions, we used another method described in Appendix~\ref{appendix:b}, where we found that $\nu_{\vec{v}}=\nu_{\rm opt}$,
$\nu_{v^{2}}=\nu_{\rm opt}(2/(1+\alpha))(f_{D}/f_{v^{2}})$, and
got the following
\begin{multline}
\label{eq:GDBB2}
\Gamma_{\rm eff}^{BB}+i\Delta_{\rm eff}^{BB}=
\Gamma_{0}+i\Delta_{0}
+\frac{\omega_{D}^{2}}{2\nu_{\rm opt}}+\\
\left(
-\frac{3\Gamma_{2}^{2}}{2\nu_{\rm opt}}
+\frac{3\Delta_{2}^{2}}{2\nu_{\rm opt}}
-i\frac{3\Gamma_{2}\Delta_{2}}{\nu_{\rm opt}}
\right)
\frac{1+\alpha}{2}
\frac{f_{v^{2}}}{f_{D}}.
\end{multline}
$\Gamma_{\rm eff}^{BB}$ and $i\Delta_{\rm eff}^{BB}$ are the effective width and shift of the Lorentzian, which the quadratic speed-dependent billiard ball profile (SDBBP) \cite{Ciurylo2002BB} collapses to, under frequent velocity-changing collisions.
This expression is a generalization of Eq.~(\ref{eq:GDHC})~\cite{Stolarczyk2022} to the arbitrary mass ratio $\alpha$ case. It should be noted that the factor $f_{v^{2}}/f_{D}$ (except $\alpha=0$ and $\alpha=\infty$ for which is equal to one), is slightly greater than unity but not more than 2.36\% in the worst case of $\alpha=2$, see Table~\ref{tab:fd}.
\par
Equations~(\ref{eq:GDHC}) and~(\ref{eq:GDBB2}) become equivalent when $\alpha=1$ and corresponding $f_{v^{2}}/f_{D}=1.018756$ is approximated by unity. Some equivalence of the HC model and BB model with $\alpha=1$ was already discussed in Ref.~\cite{Ciurylo2002BB,Ciurylo2002H2,Wcislo2014}. For $\alpha=0$ and the corresponding $f_{v^{2}}/f_{D}=1$, it means in case of SC \cite{Galatry1961,Chandrasekhar1943}, Eq.~(\ref{eq:GDBB2}) provides the effective width and shift of Lorentzian to which quadratic speed-dependent Galatry profile~\cite{Ciurylo2001SDGP,Ciurylo2002BB} collapses under frequent velocity-changing collisions. In this case Eq. (\ref{eq:GDBB2}) is reduced to Eq. (\ref{eq:GDSC}).  It is worth mentioning that comparing the asymptotic behavior of hard and soft collision models we can see that the components of the effective width and shift related to suppressed Doppler effect~\cite{Dicke1953} and thermally-averaged collisional broadening and shift are the same for both models. In contrast, the components related to the speed dependence of collisional broadening and shift are two times smaller in the case of the soft collision model. It is a natural consequence of the fact that $\nu_{v^{2}}^{SC}=2\nu_{v^{2}}^{HC}$, when we keep the same $\nu_{\vec{v}}^{SC}=\nu_{\vec{v}}^{HC}$ for both models.

\section{Components of effective width and shift}
\label{sec:components}

In line with our previous paper~\cite{Stolarczyk2022}, it is important to note that the Dicke parameter can take on complex values, $\nu_{\rm opt}=\nu_{\rm opt}^{r}+i\nu_{\rm opt}^{i}$~\cite{Rautian1967, Hess1972, Demeio1995, Pine1994}. Therefore, we can decompose the effective Lorentzian width and shift of the asymptotic quadratic correlated speed-dependent billiard ball profile~\cite{Ciurylo2002BB, Ciurylo2002icsls, Lisak2006, Wcislo2018} into several contributions:
\begin{subequations}
\begin{equation}
\label{eq:Geffi}
\Gamma_{\rm eff}=\Gamma_0+\Gamma_\gamma+\Gamma_\delta
+\Gamma_{\gamma\delta}+\Gamma_{\omega_D},
\end{equation}
\begin{equation}
\label{eq:Deffi}
\Delta_{\rm eff}=\Delta_0+\Delta_\gamma+\Delta_\delta
+\Delta_{\gamma\delta}+\Delta_{\omega_D}.
\end{equation}
\end{subequations}
Assuming quadratic speed-dependent collisional broadening and shift as well as velocity-changing collisions described by the billiard ball model, these contributions are:
\begin{subequations}
\begin{equation}
\label{eq:GGi}
\Gamma_{\gamma}=-\frac{f_{v^{2}}}{f_{D}}\frac{1+\alpha}{2}
\frac{\nu_{\rm opt}^{r}}{|\nu_{\rm opt}|^2}
\frac{3}{2}\Gamma_{2}^{2},
\end{equation}
\begin{equation}
\label{eq:GDi}
\Gamma_{\delta}=\frac{f_{v^{2}}}{f_{D}}\frac{1+\alpha}{2}
\frac{\nu_{\rm opt}^{r}}{|\nu_{\rm opt}|^2}
\frac{3}{2}\Delta_{2}^{2},
\end{equation}
\begin{equation}
\label{eq:GGDi}
\Gamma_{\gamma\delta}=-2
\frac{f_{v^{2}}}{f_{D}}\frac{1+\alpha}{2}
\frac{\nu_{\rm opt}^{i}}{|\nu_{\rm opt}|^2}
\frac{3}{2}\Gamma_{2}\Delta_{2},
\end{equation}
\begin{equation}
\label{eq:GOi}
\Gamma_{\omega_D}=\frac{\nu_{\rm opt}^{r}}{|\nu_{\rm opt}|^2}
\frac{1}{2}\omega_{D}^2,
\end{equation}
\begin{equation}
\label{eq:DGi}
\Delta_{\gamma}=\frac{f_{v^{2}}}{f_{D}}\frac{1+\alpha}{2}
\frac{\nu_{\rm opt}^{i}}{|\nu_{\rm opt}|^2}
\frac{3}{2}\Gamma_{2}^{2},
\end{equation}
\begin{equation}
\label{eq:DDi}
\Delta_{\delta}=-\frac{f_{v^{2}}}{f_{D}}\frac{1+\alpha}{2}
\frac{\nu_{\rm opt}^{i}}{|\nu_{\rm opt}|^2}
\frac{3}{2}\Delta_{2}^{2},
\end{equation}
\begin{equation}
\label{eq:DGDi}
\Delta_{\gamma\delta}=-2
\frac{f_{v^{2}}}{f_{D}}\frac{1+\alpha}{2}
\frac{\nu_{\rm opt}^{r}}{|\nu_{\rm opt}|^2}
\frac{3}{2}\Gamma_{2}\Delta_{2},
\end{equation}
\begin{equation}
\label{eq:DOi}
\Delta_{\omega_D}=-\frac{\nu_{\rm opt}^{i}}
{|\nu_{\rm opt}|^2}
\frac{1}{2} \omega_{D}^2.
\end{equation}
\end{subequations}
For real $\nu_{\rm opt}=\nu_{\rm opt}^{r}$ ($\nu_{\rm opt}^{i}=0$) the components of effective width are: $\Gamma_{\omega_{D}}$ $\Gamma_{0}$, $\Gamma_{\gamma}$, $\Gamma_{\delta}$. The first two terms represent collisionally-suppressed Doppler broadening and velocity-averaged collisional width, respectively. $\Gamma_{\gamma}$ can be seen as a reduction of line width caused by the speed-dependence of collisional broadening. It is qualitatively coherent with findings in the other context of the speed-dependent Voigt profile (SDVP)~\cite{Berman1972,Ward1974}, where also the narrowing of the spectral line was triggered by speed-dependence of collisional broadening. On the other hand, $\Gamma_{\delta}$ represents an additional broadening of the line caused by the speed-dependent spread of collisional shift~\cite{Pickett1980,Farrow1989,Wcislo2015}. The character of these contributions does not depend on the sign of $\Gamma_{2}$ and $\Delta_{2}$ parameters. The effective shift, in such circumstances, has only two components, $\Delta_{0}$ and $\Delta_{\gamma\delta}$. The first one represents an ordinary velocity-averaged collisional shift, however, the second term is less obvious and is related to the product of speed dependencies of collisional broadening and shift or their correlation~\cite{Stolarczyk2022}.
\par
The other contributions, namely $\Gamma_{\gamma\delta}$, $\Delta_{\gamma}$, $\Delta_{\delta}$, and $\Delta_{\omega_{D}}$, only come into play if $\nu_{\rm opt}$ has a nonzero imaginary component, i.e., $\nu_{\rm opt}^{i}\neq 0$. While we will not delve into a detailed discussion of these contributions, we will touch on the last one, $\Delta_{\omega_{D}}$. This shift arises due to Dicke-suppressed Doppler broadening, and it should decrease inversely with pressure, similar to the well-known $\Gamma_{\omega_{D}}$~\cite{Dicke1953}. This is because $\nu_{\rm opt}$ is proportional to pressure, while $\omega_{D}$ remains constant at a given temperature. Therefore, in a moderate range of pressures, this term can potentially impact the precise determination of the line position. In fact, as we demonstrate in the following section, under certain conditions, $\Delta_{\omega_{D}}$ can contribute up to 2 MHz to the effective line shift, $\Delta_{\rm eff}$.

To describe the relative contribution of each effect ($x=\gamma,\;\delta,\;\gamma\delta\;{\rm or}\; \omega_{D}$) under specific physical conditions, we introduce dimensionless parameters $\Gamma_{x}/\Gamma_{0}$ and $\Delta_{x}/\Delta_{0}$. It is worth noting that all collisional parameters, including $\nu_{\rm opt}$, $\Gamma_{0}$, $\Gamma_{2}$, $\Delta_{0}$, and $\Delta_{2}$, are proportional to gas pressure. Therefore, the dimensionless parameters $\Gamma_{x}/\Gamma_{0}$ and $\Delta_{x}/\Delta_{0}$ for $x=\gamma,\;\delta,\;{\rm and}\;\gamma\delta$ are independent of pressure. In fact, as we show in the following section, under certain conditions, $\Gamma_{\delta}$ can account for approximately 35\% of $\Gamma_{0}$, and the absolute value of $\Delta_{\gamma\delta}$ can reach 5\% of $\Delta_{0}$ or 20\% of $\Gamma_{0}$. For further details, please refer to the next section.

\section{Numerical validation}
\label{sec:validation}

\begin{figure}[]
    \centering
    \includegraphics[width=0.5\textwidth]{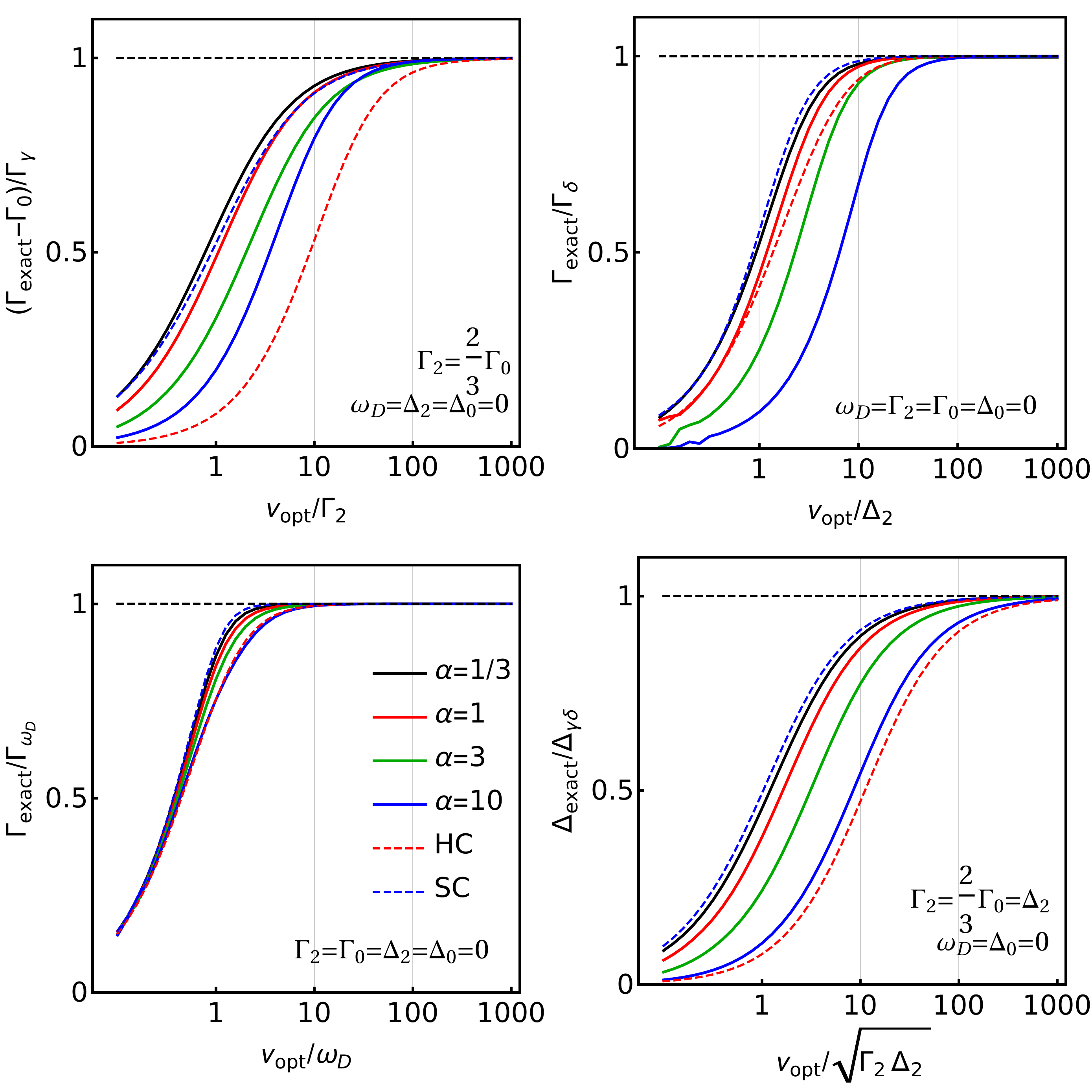}
    \caption{Numerical verification of asymptotic formulas for $\Gamma_{\gamma}$, $\Gamma_{\delta}$, $\Gamma_{\omega_D}$, and $\Delta_{\gamma\delta}$. Panels (a)-(c) show ratios of $\Gamma_{\rm exact}/\Gamma_{\gamma}$, $\Gamma_{\rm exact}/\Gamma_{\delta}$, $\Gamma_{\rm exact}/\Gamma_{\omega_D}$ as functions of $\nu_{\rm opt}/\Gamma_{2}$, $\nu_{\rm opt}/\Delta_{2}$, $\nu_{\rm opt}/\omega_{D}$, respectively, where $\Gamma_{\rm exact}$ is numerically evaluated half width at half maximum (HWHM) of the quadratic speed-dependent billiard ball profile. Panel (d) shows the ratio of $\Delta_{\rm exact}/\Delta_{\gamma\delta}$ as a function of $\nu_{\rm opt}/\sqrt{\Gamma_{2}\Delta_{2}}$ where $\Delta_{\rm exact}$ is numerically evaluated frequency corresponding to a maximum of the quadratic speed-dependent billiard ball profile. Presented ratios were evaluated for $\alpha=1/3,\; 1,\; 3,\; 10$ as well as for $\alpha=0$ it means soft collisions (SC) and hard collisions (HC).}
    \label{fig:nu}
\end{figure}

We validated the analytical expressions obtained in the previous section. To achieve this, we compared our results with the numerical half width at half maximum (HWHM) $\Gamma_{\rm exact}$ and the frequency $\Delta_{\rm exact}$ corresponding to the position of the maximum of the quadratic speed-dependent billiard ball profile~\cite{Ciurylo2002BB}. Specifically, we focused on validating the formulas for $\Gamma_{\gamma}$, $\Gamma_{\delta}$, $\Gamma_{\omega_D}$, and $\Delta_{\gamma\delta}$ given by Eqs.~(\ref{eq:GGi}), (\ref{eq:GDi}), (\ref{eq:GOi}), and (\ref{eq:DGDi}), respectively. To carry out our calculations, we considered four values of $\alpha=1/3, 1, 3, 10$, assuming real $\nu_{\rm opt}=\nu_{\rm opt}^{r}$, $\nu_{\rm opt}^{i}=0$.
\par
To provide further insight, we also compared our results with those obtained from the soft collision model, which corresponds to the billiard ball model with $\alpha=0$ and the speed-dependent Galatry profile \cite{Ciurylo2001SDGP} and the hard collision model \cite{Nelkin1964,Rautian1967} which we discuss in detail in Ref.~\cite{Stolarczyk2022}. To simplify our analysis, we focused on one effect at a time and set the appropriate values for the line shape parameters: $\Gamma_{0}$, $\Delta_{0}$, $\Gamma_{2}$, $\Delta_{2}$, $\omega_{D}$, $\nu_{\rm opt}$ in the quadratic SDBBP.
\par
Our results, shown in Fig.~\ref{fig:nu}, demonstrate that $\Gamma_{\rm exact}$ converges to $\Gamma_{\gamma}$, $\Gamma_{\delta}$, $\Gamma_{\omega_D}$ as well as $\Delta_{\rm exact}$ approaches $\Delta_{\gamma\delta}$ in the limit of high $\nu_{\rm opt}$. However, we observed that the convergence is slower for higher values of $\alpha$ in the case of $\Gamma_{\gamma}$, $\Gamma_{\delta}$, and $\Delta_{\gamma\delta}$. This observation is directly related to the fact that $\nu_{v^{2}}$ decreases with increasing $\alpha$, while $\nu_{\vec{v}}$ remains constant. On the other hand, the situation is different for $\Gamma_{\omega_D}$, which is determined by $\nu_{\vec{v}}$ and is not dependent on $\alpha$. As shown in Fig.\ref{fig:nu}(c), the variations for $\alpha$ are weak in this case. However, for extremely large values of $\alpha$, significant discrepancies for different $\alpha$ are observed. This topic has been discussed in detail in Ref.~\cite{Ciurylo2002BB}.

\begin{figure}[b]
    \centering
    \includegraphics[width=0.5\textwidth]{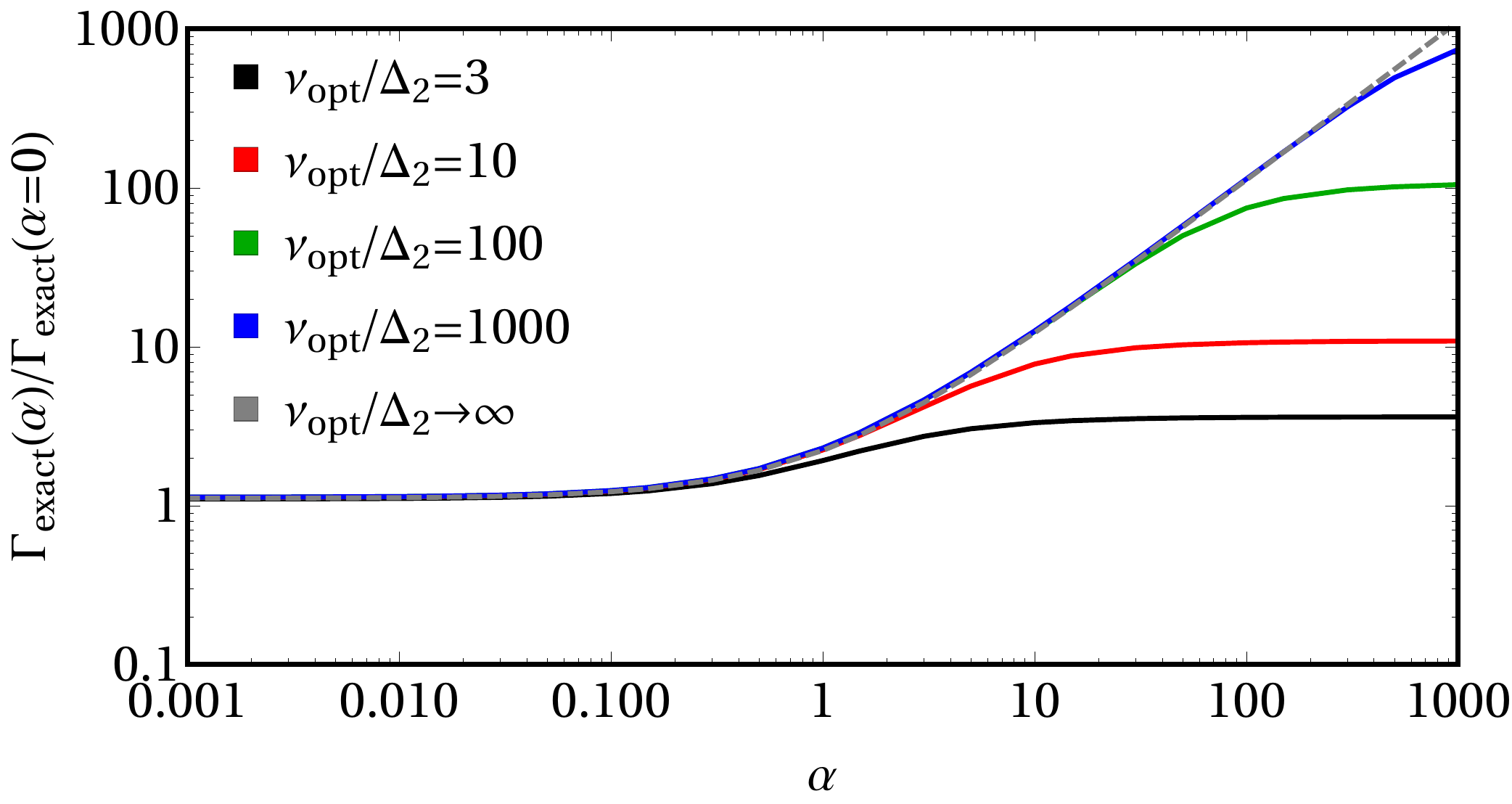}
    \caption{Ratio of $\Gamma_{\rm exact}(\alpha)/\Gamma_{\rm exact}(\alpha=0)$ evaluated for several values $\nu_{opt}/\Delta_2$ within quadratic speed-dependent billiard ball profile in which parameters $\Gamma_0$, $\Delta_0$, $\Gamma_2$, $\nu_{\rm opt}^i$, and $\omega_{D}$ were set to zero and its comparison with ratio $\Gamma_{\delta}(\alpha)/\Gamma_{\delta}(\alpha=0)\approx 1+\alpha$ obtained for $\nu_{\rm opt}/\Delta_2\rightarrow\infty$.}
    \label{fig:alpha}
\end{figure}

The discussion of the effective broadening component $\Gamma_{\delta}$, which arises from the speed dependence of the collisional shift $\Delta(v)$, is important in this context. As mentioned earlier, an increase in $\alpha$ results in a decrease in $\nu_{v^2}$. Therefore, we anticipate that, with other parameters held constant, an increase in $\alpha$ will result in an increase in $\Gamma_{\delta}$. In the frequent velocity-changing collisions regime, where $\nu_{\rm opt}$ dominates, an asymptotic analytical relation for quadratic SDBBP~\cite{Ciurylo2002BB} can be written as:
\begin{equation}
\label{eq:ratio}
\frac{\Gamma_{\delta}(\alpha)}{\Gamma_{\delta}(\alpha=0)}=\frac{f_{v^{2}}}{f_{D}}(1+\alpha)\approx1+\alpha.
\end{equation}
To simplify the calculation, the factor $f_{v^2}/f_D$ in the asymptotic relation can be approximated as unity, introducing an error of less than 2\%. We verified the accuracy of this approximation for finite values of $\nu_{\rm opt}/\Delta_2$ by simulating quadratic SDBBP for various values of $\nu_{\rm opt}/\Delta_2$ and $\alpha$, while setting all other parameters ($\Gamma_0$, $\Delta_0$, $\Gamma_2$, $\nu_{\rm opt}^i$, and $\omega_D$) to zero. We then calculated $\Gamma_{\rm exact}$ for the simulated profiles, and plotted the ratios $\Gamma_{\rm exact}(\alpha)/\Gamma_{\rm exact}(\alpha=0)$ as a function of $\alpha$ for several values of $\nu_{\rm opt}/\Delta_2$ (3, 10, 100, 1000) in Fig.\ref{fig:alpha}. As shown in this figure, for $\nu_{\rm opt}/\Delta_2=10$, the simulation results are reproduced by Eq.(\ref{eq:ratio}) for small to moderate values of $\alpha$ up to $\alpha=3$.

\begin{table}[htbp]
\centering
\caption{Comparison of the Lorentz profile effective parameters for Q(1)~1-0 transition in H$_2$ perturbed by He ($\alpha=2$) at 10~atm and 296~$K$ obtained with Eqs.~\eqref{eq:GGi}-\eqref{eq:DOi} derived from the billiard ball model (BB) and effective parameters from Ref.~\cite{Stolarczyk2022,Wcislo2021_database} derived from hard collision model (HC). The line-shape parameters are $\Gamma_0 = 22.346$, $\Delta_0 = 90.971$, $\Gamma_2 = 10.856$, $\Delta_2 = 38.192$, $\nu_{\rm opt}^r = 404.148$, $\nu_{\rm opt}^i = -54.697$, $\omega_D = 21.659$~\cite{Wcislo2021_database} are given in the units of $10^{-3}\;{\rm cm}^{-1}$.}
\begin{tabular}{lcccc}
\hline
\hline
$\Gamma_{x}$& $\Gamma^{BB}_{x}(10^{-3}\;{\rm cm}^{-1})$ & $\Gamma^{BB}_{x}/\Gamma_{0}$ & $\Gamma^{HC}_{x}(10^{-3}\;{\rm cm}^{-1})$ & $\Gamma^{HC}_{x}/\Gamma_{0}$ \\
\hline
$\Gamma_0$ & 22.346 & 1.000 & 22.346 & 1.000\\
$\Gamma_\gamma$ & -0.645 & -0.029 & -0.430 & -0.019\\
$\Gamma_\delta$ & 7.974 & 0.357 & 5.316 & 0.238 \\
$\Gamma_{\gamma\delta}$ & 0.613 & 0.027 & 0.409 & 0.018 \\
$\Gamma_{\omega_D}$ & 0.572 & 0.026 & 0.572 & 0.026\\
\hline
$\Gamma_{\rm eff}$ & 30.860 & 1.381 & 28.215 & 1.263 \\

\textcolor{\kolora}{$\Gamma_{\rm exact}$} & \textcolor{\kolora}{30.140} & \textcolor{\kolora}{1.349} & \textcolor{\kolora}{27.388} & \textcolor{\kolora}{1.226} \\
\hline
\hline$\Delta_{x}$& $\Delta^{BB}_{x}(10^{-3}\;{\rm cm}^{-1})$ & $\Delta^{BB}_{x}/\Delta_{0}$ & $\Delta^{HC}_{x}(10^{-3}\;{\rm cm}^{-1})$ & $\Delta^{HC}_{x}/\Delta_{0}$ \\
\hline
$\Delta_0$ & 90.971 & 1.000 & 90.971 & 1.000\\
$\Delta_\gamma$ & -0.087 & 0.001 & -0.058 & -0.0006\\
$\Delta_\delta$ & 1.079 & 0.012 & 0.719 & 0.008\\
$\Delta_{\gamma\delta}$ & -4.533 & -0.050 & -3.022 & -0.033\\
$\Delta_{\omega_D}$ & 0.077 & 0.0008 & 0.077 & 0.0008\\
\hline
$\Delta_{\rm eff}$ & 87.507 & 0.962 & 88.688 & 0.975 \\
\textcolor{\kolora}{$\Delta_{\rm exact}$} & \textcolor{\kolora}{86.270} & \textcolor{\kolora}{0.948} & \textcolor{\kolora}{88.005} & \textcolor{\kolora}{0.967} \\
\hline
\hline

\end{tabular}
\label{tab:params}
\end{table}

\begin{figure}[b]
    \centering
    \includegraphics[width=0.5\textwidth]{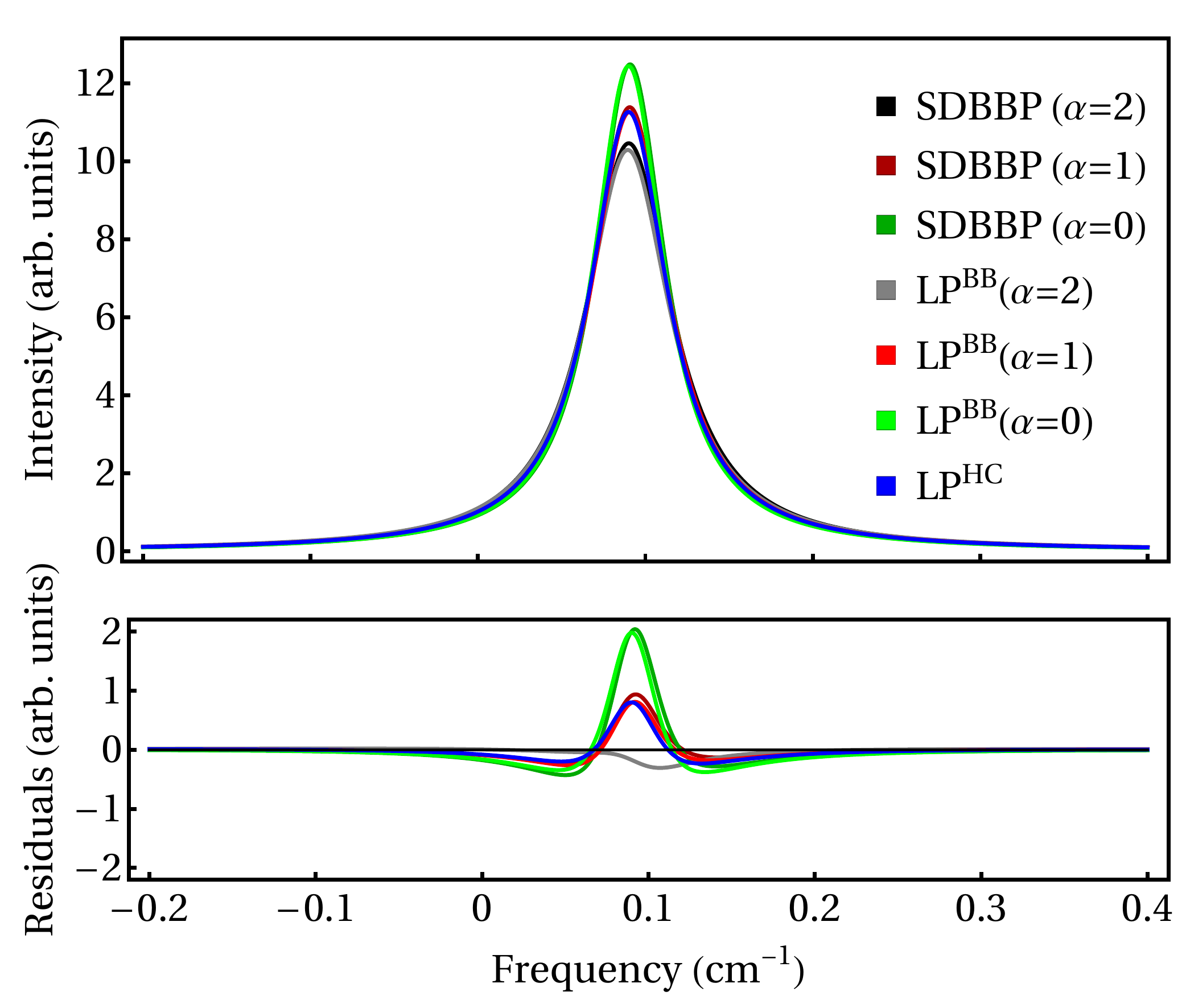}
    \caption{A comparison of the reference speed-dependent billiard ball  profiles (SDBBP) calculated with $\alpha=2,\; 1,\; 0$ and parameters listed in the caption of Tab.~\ref{tab:params} for Q(1)~1-0 transition in H$_2$ perturbed by He at 10~atm and 296~$K$ \cite{Stolarczyk2022,Wcislo2021_database} and Lorentz profiles (LP$^{BB}$) derived as the collapsed SDBBP calculated with parameters  given in this work by Eqs.~\eqref{eq:Geffi}-\eqref{eq:DOi} as well as Lorentz profile (LP$^{HC}$) derived as the collapsed hard collision profile from Ref.~\cite{Stolarczyk2022}. The area under each line is normalized to 1. The lower graph presents the residuals against reference SDBBP ($\alpha=2$) having $\alpha$ corresponding to H$_{2}$-He system with the same color notation as in the upper panel.}
    \label{fig:profile}
\end{figure}
\par
{\color{\kolora} As mentioned in Section~\ref{sec:derivation}, our derivation was conducted under the assumption that velocity-changing collisions dominate over both Doppler broadening and collisional broadening and shift, particularly their speed-dependent parts. To satisfy this condition, the interactions between the absorber and perturber molecules at the initial and final molecular states need to be either identical or closely similar~\cite{Rautian1967, Nienhuis1978, Looney1987, Ciurylo2001JQSRT, Hess1972, Kochanov1977, Nienhuis1978, Berman1982, Berman1984, Monchick1986, Thibault2017}. Physically, this requirement can be met in the case of rovibronic transitions in molecular hydrogen. This can be observed through \textit{ab initio} calculations, as presented in Ref.~\cite{Wcislo2021_database}. Additionally, it has been demonstrated in Refs.~\cite{Slowinski2020, Slowinski2022} that such calculations, along with the use of SDBBP~\cite{Ciurylo2002BB}, can achieve sub percent agreement with experimental data.
}
\par
{\color{\kolora} It is seen from Fig.~\ref{fig:nu} that for $\alpha\leq 3$ and $\nu_{\rm opt}/\Gamma_{2}\geq 10$ the error introduced by our approximation of $\Gamma_{\gamma}$ does not exceed 16\%. Similarly, for $\nu_{\rm opt}/\Delta_{2}\geq 10$ our approximation of $\Gamma_{\delta}$ reproduces the corresponding width with inaccuracy less than 7\%. Ultimately, if $\nu_{\rm opt}/\omega_{D}\geq 10$ the error on $\Gamma_{\omega_{D}}$ is at most 0.3\%, which is significantly smaller than in the previous cases. On the other hand, for the pressure shift, if $\nu_{\rm opt}/\sqrt{\Gamma_{2}\Delta_{2}}\geq 10$ our approximation of $\Delta_{\gamma\delta}$ introduces error, which does not exceed 23\%. In conditions where the speed-averaged values $\Gamma_{0}$ and $\Delta_{0}$ are significantly greater than their speed dependencies, $\Gamma_{2}$ and $\Delta_{2}$, the combined errors originating from these contributions result in a significantly smaller relative error on the entire line profile. Additionally, if the $\nu_{opt}$ dominates over the other line shape parameters (i.e., $\Gamma_2$, $\Delta_2$, $\omega_D$) by several dozens, our approximation should reproduce the original profile with several percent accuracy.
}
\par
We compared the collapsed Lorentz profile to the SDBBP~\cite{Ciurylo2002BB}. To do this, we examined the Q(1) 1-0 transition in H$_2$ perturbed by He at a pressure of 10 atm and a temperature of 296~K. We used the same line-shape parameters as in Ref.~\cite{Stolarczyk2022}, which were obtained from Ref.~\cite{Wcislo2021_database}. The values of the line shape parameters can be found in the caption of Tab.~\ref{tab:params}. This table lists all contributions $\Gamma_{x}$ and $\Delta_{x}$ to the effective Lorentzian width $\Gamma_{\rm eff}$ and shift $\Delta_{\rm eff}$, respectively. These values were calculated using Eqs.~\eqref{eq:GGi}-\eqref{eq:DOi} under the assumption of a billiard ball model ($\alpha=2$) for velocity-changing collisions and the hard collision model, as described in Ref.~\cite{Stolarczyk2022}. {\color{\kolora} In Tab.~\ref{tab:params} and Fig.~\ref{fig:profile}, we used the unit of wave numbers $\widetilde{\nu}=2\pi\omega/c$ instead of circular frequencies $\omega$, which is more common in molecular spectroscopy for line shape parameters.}
\par
The ratios $\nu^{r}_{\rm opt}/\omega_D=18.7$, $\nu^{r}_{\rm opt}/\Gamma_{2}=37.2$, and $\nu^{r}_{\rm opt}/\Delta_{2}=10.6$ demonstrate that our assumption from Sec.~\ref{sec:derivation}, that the velocity-changing collisions dominate over Doppler broadening and collisional broadening and shift, is valid for the chosen conditions (pressure, temperature, molecular system). It is important to note that the values of $\Gamma_{0}$ and $\Delta_{0}$ are not relevant in this context, as they only result in a simple convolution with the Lorentz profile determined by these parameters, which are additive in the case of Lorentz profiles. 
{\color{\kolora} Furthermore, we have verified that the binary collision approximation holds well under these conditions, and the contribution of three-body collisions to collisional width and shift should not exceed the percentage level. For a detailed discussion, see Section II.B3 in Ref. \cite{Hartmann2008book}.}
\par
The H$_{2}$ spectral line perturbed by He, for which $\alpha=2$, allows us to observe the importance of the perturber/absorber mass ratio for the collapse of the spectral line shape to a Lorentz profile due to frequent velocity-changing collisions. As shown in Tab.~\ref{tab:params}, the use of the BB model results in a Lorentz profile that is 9.4\% wider than the profile obtained from the HC model~\cite{Stolarczyk2022}. This is mainly due to the fact that $\Gamma_{\delta}$, the second most significant contribution to $\Gamma_{\rm eff}$ after $\Gamma_{0}$, is 50\% larger in the case of the BB model compared to the HC model. \textcolor{\kolora}{It should be noted that the $\Gamma_{\rm eff}$ and $\Delta_{\rm eff}$ obtained from Eqs.~\eqref{eq:Geffi} and~\eqref{eq:Deffi}, agree with the $\Gamma_{\rm exact}$ and $\Delta_{\rm exact}$ obtained from the original SDBBP within X and Y \%, respectively.}

Figure~\ref{fig:profile} compares the SDBBP \cite{Ciurylo2002BB} with $\alpha=2,\; 1,\; 0$ with Lorentz profiles derived from the BB model (LP$^{BB}$) and the HC model (LP$^{HC}$). It shows that LP$^{BB}$($\alpha=2$) provides a good approximation of SDBBP($\alpha=2$), with a difference of about 2\% at the maximum of the profile. In contrast, LP$^{HC}$ exhibits a much larger deviation. The LP$^{HC}$ follows closely LP$^{BB}$($\alpha=1$) and SDBBP($\alpha=1$). The discrepancy between SDBBP($\alpha=2$) and LP$^{HC}$ is caused by the fact that, in the case of the HC model, the frequency of speed-changing collisions $\nu_{v^{2}}$ does not capture its dependence on $\alpha$, which is about 2/3 times smaller for the BB model with the same parameter values and $\alpha=2$. On the other hand the frequency of speed-changing collisions $\nu_{v^{2}}$ for HC model agrees within 2\% with those from BB model with $\alpha=1$. A comparison of SDBBP($\alpha=2$) with soft collision profiles LP$^{SC}$ and SDGP \cite{Ciurylo2001SDGP} which are equivalent to LP$^{BB}$($\alpha=0$) and SDBBP($\alpha=0$), respectively, looks even worse. It is caused by the fact that the BB model with the same parameter values and $\alpha=2$ has the frequency of speed-changing collisions $\nu_{v^{2}}$ only about 1/3 of this in SC model (BB model with $\alpha=0$).

\section{Conclusions}
\label{sec:conclusions}
In summary, we demonstrated that under the limit of frequent velocity-changing collisions, the speed-dependent Dicke-narrowed profile of the spectral line collapses to the Lorentz profile. Our work provides formulas for effective Lorentzian width and shift, which take into account the arbitrary speed dependence of collisional broadening and shift, as well as the velocity-changing collision operator. Specifically, for the asymptotic behavior of the quadratic speed-dependent billiard ball profile~\cite{Ciurylo2002BB}, we obtained simple analytical expressions for Lorentzian width and shift. Our results generalize those recently reported in Ref.~\cite{Stolarczyk2022} for the speed-dependent hard collision profile~\cite{Pine1999} to the case of arbitrary perturber/absorber mass ratio, $\alpha$. We verified the applicability of our formulas by comparing them with numerical calculations of quadratic speed-dependent billiard ball profiles over a wide range of line-shape parameters.
\par
Finally, our comparison with the numerical calculation of the H$_2$ spectral line perturbed by He showed good agreement with the profile derived in this work. This is thanks to properly accounting for the perturber/absorber mass ratio, which was not possible with the hard collision model. Our results offer a more accurate and general description of spectral line shapes under the limit of frequent velocity-changing collisions, with potential applications in various fields including atmospheric and astrophysical spectroscopy {\color{\kolora} involving rovibrational transitions of molecular hydrogen}.

\section*{Acknowledgements}
N.S. was supported by Polish National Science Centre Project No. 2019/35/N/ST2/04145. 
P.W. was supported by Polish National Science Centre Project No. 2019/35/B/ST2/01118. 
R.C. was supported by Polish National Science Centre Project No. 2021/41/B/ST2/00681.
The research is a part of the program of the National Laboratory FAMO in Toru\'n, Poland.
Calculations have been partially carried out using resources provided by the Wroclaw Centre for Networking and Super-computing \url{( http://wcss.pl )}, Grant No. 546. 
We gratefully acknowledge Polish high-performance computing infrastructure PLGrid (HPC Centers: ACK Cyfronet AGH, CI TASK) for providing computer facilities and support within computational grant no. PLG/2023/016279.

\appendix
\section{The Burnett functions representation}
\label{appendix:a}
\setcounter{figure}{0}

For evaluation of the speed-dependent Dicke-narrowed spectral line shape from the transport/relaxation equation we use a subset of the Burnett functions having axial symmetry about the wave vector $\vec{k}$~\cite{Lindenfeld1979}. A detailed discussion of the application of this set of basis functions, as well as matrix representation of the operators present in the transport/relaxation equation, can be found in Ref.~\cite{Ciurylo2002BB}. For convenience we recall these results here, using the original formulation.
\par
We assume the basis functions in the following form~\cite{Lindenfeld1980}:
\begin{equation}
\label{eq:a1}
\varphi_{nl}(\vec{v})=N_{nl}\;(v/v_{m_{A}})^{l}\;
L_{n}^{l+1/2}(v^{2}/v_{m_{A}}^{2})\;
P_{l}(\vec{e}_{k} \cdot \vec{e}_{v}),
\end{equation}
where a normalization factor,
\begin{equation}
\label{eq:a2}
N_{nl}=\sqrt{\frac{\pi^{1/2}\;n!\;(2l+1)}{2\;\Gamma(n+l+3/2)}}\; ,
\end{equation}
$\Gamma(...)$ is the gamma-Euler function. The associated Laguerre polynomials are defined as
\begin{equation}
\label{eq:a3}
L_{n}^{l+1/2}(x^{2})=\sum_{m=0}^{n}
\frac{(-1)^{m}\; \Gamma(n+l+3/2)}{m!\;(n-m)!\;\Gamma(m+l+3/2)}\;x^{2m} \; ,
\end{equation}
where $x=v/v_{m_{A}}$ is the reduced speed of the active molecule. The Legendre polynomials are defined as
\begin{equation}
\label{eq:a4}
P_{l}(y)=\frac{1}{2^{l}}\sum_{k=0}^{[l/2]}
\frac{(-1)^{k}\; (2l-2k)!}{k!\;(l-k)!\;(l-2k)!}\;y^{l-2k} \; ,
\end{equation}
where $y=\vec{e}_{k} \cdot \vec{e}_{v}$ is the cosine of the angle between the velocity vector $\vec{v}=v\vec{e}_{v}$ and the wave vector $\vec{k}=k\vec{e}_{k}$, $\vec{e}_{v}$ and $\vec{e}_{k}$ are unit vectors.
\par
The basis functions, $\varphi_{nl}(\vec{v})$, are eigenfunctions of the velocity-changing collision operator in the case of the soft collisions, where perturber/absorber mas ratio $\alpha=0$. Like in~\cite{Ciurylo2002BB}, when calculating matrix elements of an operator, $\hat{A}$, we use the following notation $[\mbox{\bf A}]_{nl,n'l'}=(\varphi_{nl}(\vec{v}),\hat{A}\varphi_{n'l'}(\vec{v}))=\langle nl|\hat{A}|n'l'\rangle$.
\par
The Doppler shift operator, $\vec{k} \cdot \vec{v}=\omega_{D} \vec{e}_{k} \cdot \vec{v}/v_{m_{A}}$, is represented by a matrix, $\mbox{\bf K}$, the elements of which are:~\cite{Ciurylo2002BB}
\begin{equation}
\label{eq:a5}
[\mbox{\bf K}]_{nl,n'l'}=\omega_{D}\langle nl|\vec{e}_{k} \cdot \vec{v}/v_{m}|n'l'\rangle \; ,
\end{equation}
where $\omega_{D}=kv_{m_{A}}$ and
\begin{multline}
\label{eq:a6}
\langle nl|\vec{e}_{k} \cdot \vec{v}/v_{m}|n'l'\rangle =\\
\left[
\sqrt{n+l+3/2}\;\delta_{n,n'}
-\sqrt{n}\;\delta_{n,n'+1}
\right] \\
\times
\sqrt{\frac{(l+1)^{2}}{4(l+1)^{2}-1}}
\;\delta_{l,l'-1}
\\
+
\left[
\sqrt{n+l+1/2}\;\delta_{n,n'}
-\sqrt{n+1}\;\delta_{n,n'-1}
\right]\\
\times
\sqrt{\frac{l^{2}}{4l^{2}-1}}
\;\delta_{l,l'+1} .
\end{multline}
\par
The speed-dependent collisional width and shift operator, $\hat{S}_{D}^{f}=-\Gamma(v)-i\Delta(v)$, is represented by a matrix, $\mbox{\bf S}_{D}^{f}$, the elements of which are:~\cite{Ciurylo2002BB,Robert1998}
\begin{multline}
\label{eq:a7}
[\mbox{\bf S}_{D}^{f}]_{nl,n'l'}=\\
\frac{4}{\sqrt{\pi}(2l+1)}\;N_{nl}\;N_{n'l}\;\delta_{l,l'}
\int_{0}^{\infty}dx\; e^{-x^{2}} x^{2l+2}\;
\times
\\
\times
L_{n}^{l+1/2}(x^{2})\;L_{n'}^{l+1/2}(x^{2})\;
\left[\Gamma(xv_{m_{A}})+i\Delta(xv_{m_{A}})\right] .
\end{multline}
In the case of quadratic speed-dependent collisional broadening and shift~\cite{Rohart1994},
\begin{equation}
\label{eq:a8}
\Gamma(v)+i\Delta(v)= \Gamma_{0}+i\Delta_{0} +(\Gamma_{2}+i\Delta_{2})\left(\frac{v^{2}}{v_{m}^{2}}-\frac{3}{2}\right),
\end{equation}
the matrix elements can be written as
\begin{multline}
\label{eq:a9}
[\mbox{\bf S}_{D}^{f}]_{nl,n'l'}=
-\left[\Gamma_{0}+i\Delta_{0}-(\Gamma_{2}+i\Delta_{2})3/2\right]
\delta_{n,n'}\delta_{l,l'}
\\
-\left[\Gamma_{2}+i\Delta_{2}\right]
\langle nl|v^{2}/v_{m}^{2}|n'l'\rangle ,
\end{multline}
where 
\begin{multline}
\langle nl|v^{2}/v_{m}^{2}|n'l'\rangle=\\
\left[
(2n+l+3/2)\;\delta_{n,n'}
-\sqrt{(n+l+1/2)n}\;\delta_{n,n'+1}
\right.
\\
\left.
-\sqrt{(n+l+3/2)(n+1)}\;\delta_{n,n'-1}
\right]\delta_{l,l'} .
\label{e10}
\end{multline}
\par
The velocity-changing collisions operator $\hat{S}_{BB}^{f}$ in the billiard ball model is represented by a matrix, $\mbox{\bf S}_{BB}^{f}$, the elements of which are:~\cite{Lindenfeld1979,Lindenfeld1980}
\begin{equation}
\label{eq:a11}
[\mbox{\bf S}_{BB}^{f}]_{nl,n'l'}=\nu^{(0)}\; M_{nl,n'l'}^{E*} ,
\end{equation}
where $\nu^{(0)}= v_{m}^{2}/(2D^{(0)})$ and
\begin{equation}
\label{eq:a12}
D^{(0)}=\frac{3}{8} \left(\frac{k_{B}T}{2\pi\mu}\right)^{1/2}
\frac{1}{N\sigma^{2}}
\end{equation}
is the first-order self-diffusion coefficient for rigid spheres, $\sigma$ is the average of the rigid sphere diameter of the absorber and perturber, $N$ is the number density of perturbers, $\mu=m_{A}m_{P}/(m_{A}+m_{P})$ is the reduced mass. The analytical expressions for coefficients $M_{nl,n'l'}^{E*}$ for billiard ball model were derived by Lindenfeld and Shizgal~\cite{Lindenfeld1979,Lindenfeld1980}
\begin{multline}
\label{eq:a13}
M_{nl,n'l'}^{E*}=\\
-\delta_{l,l'}\;\frac{3\;l!}{8 M_{2}}
\sqrt{\frac{n!\;n'!}{\Gamma(n+l+3/2)\Gamma(n'+l+3/2)}}
\times
\\
\Bigg\{
\sum_{p=0}^{\tilde{n}}\;
\sum_{s=0}^{\tilde{n}-p}\;
\sum_{m=0}^{\tilde{n}-p-s}\;
\sum_{q=0}^{l}\;
\sum_{r=0}^{l-q}\;
\left[
\frac{4^{p}\;(r+s+p+q+1)!}
{(p+q+1)!\;r!\;s!}
\right]
\times
\\
\left[
\frac{\Gamma(n+n'-2s-2p-m+l-r-q-1/2)\;B_{p,q}^{(1)}(\infty)}
{(n-m-s-p)!\;(n'-m-s-p)!\;(l-r-q)!\;m!}
\right]
\times
\\
\left[
M_{1}^{l+p-r-q}M_{2}^{n+n'+q-2m-2s-p}(M_{1}-M_{2})^{m+r+2s}
\right]
\Bigg\} ,
\end{multline}
where $M_{1}=m_{A}/(m_{A}+m_{P})=1-M_{2}$,
$\tilde{n}={\rm min}(n,n')$ and
\begin{equation}
\label{eq:a14}
B_{p,q}^{(1)}(\infty)=\frac{(2p+q+1)!}{2q!\;(2p+1)!}
-\frac{2^{q-1}(p+q+1)!}{p!\;q!} \; .
\end{equation}
The exact diffusion coefficient $D$ differs from $D^{(0)}$ and they are related by coefficient $f_{D}=D/D^{(0)}$ which evaluation is explained in Appendix~\ref{appendix:b}. Using these quantities the matrix elements can be written as
\begin{equation}
\label{eq:a15}
[\mbox{\bf S}_{BB}^{f}]_{nl,n'l'}=\nu_{\rm diff}\; f_{D}\; M_{nl,n'l'}^{E*}
\end{equation}
where 
\begin{equation}
\label{eq:a16}
\nu_{\rm diff} = \frac{v_{m}^{2}}{2D} 
\end{equation}
is the effective frequency of velocity-changing collisions.

\section{Non-diagonal corrections to effective Lorentzian width and shift}
\label{appendix:b}
\setcounter{figure}{0}
Aiming at calculations of spectral line shape with quadratic speed-dependent collisional broadening and shift in the limit dominated by velocity-changing collisions, Eq.~(\ref{eq:EqL}) can be rewritten in the Burnett functions representation,
\begin{subequations}
\begin{multline}
\label{eq:b1a}
1=\left(-i(\omega-\omega_{0})-[\mbox{\bf S}_{D}^{f}]_{00,00}\right)c_{00}(\omega)\\
+i[\mbox{\bf K}]_{00,01}c_{01}(\omega)
-[\mbox{\bf S}_{D}^{f}]_{00,10}c_{10}(\omega),
\end{multline}
\begin{equation}
\label{eq:b1b}
0=i[\mbox{\bf K}]_{01,00}\delta_{n,0}c_{00}(\omega)
-\sum_{n'=0}^{\infty}[\mbox{\bf S}_{VC}^{f}]_{n1,n'1}
c_{n'1}(\omega),
\end{equation}
\begin{equation}
\label{eq:b1c}
0=-[\mbox{\bf S}_{D}^{f}]_{10,00}\delta_{n,1}c_{00}(\omega)
-\sum_{n'=1}^{\infty}[\mbox{\bf S}_{VC}^{f}]_{n0,n'0}
c_{n'0}(\omega).
\end{equation}
\end{subequations}
We also take advantage from the fact that $c_{00}(\omega)$ is not coupled to other coefficients by matrix $\mbox{\bf S}_{VC}^{f}$ and it dominates two other matrices $\mbox{\bf S}_{D}^{f}$ and $\mbox{\bf K}$. Therefore, in the limit of frequent velocity-changing collisions, we are allowed to use matrix $\mbox{\bf S}_{VC}^{f}$ and only those matrix elements of $\mbox{\bf S}_{D}^{f}$ and $\mbox{\bf K}$ which provide coupling of $c_{00}(\omega)$ to other coefficients in the transport/relaxation equation.
With the velocity-changing collision operator in the form from Eq.~(\ref{eq:a11}), we can find $c_{00}(\omega)$ by solving two other sets of linear equations, 
\begin{subequations}
\begin{equation}
\label{eq:b2a}
\delta_{n,0}=-\sum_{n'=0}^{\infty}M^{E*}_{n1,n'1}a_{n'1},
\end{equation}
\begin{equation}
\label{eq:b2b}
\delta_{n,1}=-\sum_{n'=1}^{\infty}M^{E*}_{n0,n'0}a_{n'0}.
\end{equation}
\end{subequations}
Equations~(\ref{eq:b2a}) and~(\ref{eq:b2b}) are equivalent to Eqs.~(\ref{eq:b1b}) and~(\ref{eq:b1c}) when we set
\begin{subequations}
\begin{equation}
\label{eq:b3a}
c_{n1}(\omega)=
-\frac{i[\mbox{\bf K}]_{01,00}}{\nu^{(0)}}
c_{00}(\omega)a_{n1},
\end{equation}
\begin{equation}
\label{eq:b3b}
c_{n0}(\omega)=
\frac{[\mbox{\bf S}_{D}^{f}]_{10,00}}{\nu^{(0)}}
c_{00}(\omega)a_{n0}.
\end{equation}
\end{subequations}
Now we can rewrite Eq.~(\ref{eq:b1a}) in the following form,
\begin{multline}
\label{eq:b4}
1=\bigg(
-i(\omega-\omega_{0})-[\mbox{\bf S}_{D}^{f}]_{00,00}
\bigg.\\
\bigg.
+\frac{[\mbox{\bf K}]_{01,00}^{2}}{\nu^{(0)}}a_{01}
-\frac{[\mbox{\bf S}_{D}^{f}]_{10,00}^{2}}{\nu^{(0)}}a_{10}.
\bigg)c_{00}(\omega).
\end{multline}
It is convenient to introduce the coefficient $f_{D}=a_{01}(-M^{E*}_{01,01})=a_{01}$, were $a_{01}$ can be found by solving Eq.~(\ref{eq:b2a})~\cite{Lindenfeld1980}. Analogically, we can introduce $f_{v^{2}}=a_{10}(-M^{E*}_{10,10})=a_{10}2/(1+\alpha)$, were $a_{01}$ can be found by solving Eq.~(\ref{eq:b2b}).
\par
Table~\ref{tab:fd} collects the coefficients $f_{D}$ and $f_{v^{2}}$ as well as their ratios for the same set of $\alpha$. We repeat the calculations of $f_{D}$ performed by Lindenfeld in 1980~\cite{Lindenfeld1980} for the billiard ball model but with higher numerical precision (our results are in good agreement for all but the last digit of the Table~1 of Ref.\cite{Lindenfeld1980}). With these coefficients we define effective rates $\nu_{\vec{v}}=\nu^{(0)}/f_{D}$ of change of velocity vector $\vec{v}$ and $\nu_{v^{2}}=\nu^{(0)}(2/(1+\alpha))/f_{v^{2}}$ of change of $v^{2}$. From that we can obtain the relation $\nu_{v^{2}}=\nu_{\vec{v}}(2/(1+\alpha))(f_{D}/f_{v^{2}})$ between rates of change of $\vec{v}$ and $v^{2}$. 
\par
Solving Eq.(\ref{eq:b4}) yields
\begin{equation}
\label{eq:b5}
c_{00}(\omega)=\frac{1}
{\Gamma_{\rm eff}+i\Delta_{\rm eff}-i(\omega-\omega_{0})},
\end{equation}
where the effective Lorentzian width and shift are
\begin{equation}
\label{eq:b6}
\Gamma_{\rm eff}+i\Delta_{\rm eff}=
-[\mbox{\bf S}_{D}^{f}]_{00,00}
+\frac{[\mbox{\bf K}]_{01,00}^{2}}{\nu_{\vec{v}}}
-\frac{[\mbox{\bf S}_{D}^{f}]_{10,00}^{2}}{\nu_{v^{2}}} \; .
\end{equation}
This equation can be also written in the following form,
\begin{equation}
\label{eq:b7}
\Gamma_{\rm eff}+i\Delta_{\rm eff}=
\Gamma_{0}+i\Delta_{0}
+\frac{1}{2}\frac{\omega_{D}}{\nu_{\vec{v}}}
-\frac{3}{2}\frac{(\Gamma_{2}+i\Delta_{2})^{2}}{\nu_{\vec{v}}}
\frac{1+\alpha}{2}\frac{f_{v^{2}}}{f_{D}},
\end{equation}
which gives Eq.~(\ref{eq:GDBB2}).

To take into account any arbitrary speed-dependence of collisional broadening, $\Gamma(v)$, and shift, $\Delta(v)$, we need to include other matrix elements, $[\mbox{\bf S}_{D}^{f}]_{n0,00}$, besides $[\mbox{\bf S}_{D}^{f}]_{10,00}$.
In such case, Eqs.~(\ref{eq:b1a})-(\ref{eq:b1b}) should be generalized to the following form,
\begin{subequations}
\begin{multline}
\label{eq:b8a}
1=\left(-i(\omega-\omega_{0})-[\mbox{\bf S}_{D}^{f}]_{00,00}\right)c_{00}(\omega)\\
+i[\mbox{\bf K}]_{00,01}c_{01}(\omega)
-\sum_{n'=1}^{\infty}
[\mbox{\bf S}_{D}^{f}]_{00,n'0}c_{n'0}(\omega),
\end{multline}
\begin{equation}
\label{eq:81b}
0=i[\mbox{\bf K}]_{01,00}\delta_{n,0}c_{00}(\omega)
-\sum_{n'=0}^{\infty}[\mbox{\bf S}_{VC}^{f}]_{n1,n'1}
c_{n'1}(\omega),
\end{equation}
\begin{equation}
\label{eq:b8c}
0=-[\mbox{\bf S}_{D}^{f}]_{n0,00}c_{00}(\omega)
-\sum_{n'=1}^{\infty}[\mbox{\bf S}_{VC}^{f}]_{n0,n'0}
c_{n'0}(\omega).
\end{equation}
\end{subequations}
The coefficient $c_{00}(\omega)$can be found in analogical way like in Eqs.~(\ref{eq:b1a})-(\ref{eq:b1b}). The only difference is that the set of linear equations, Eq.~(\ref{eq:b2b}), is replaced by
\begin{equation}
\label{eq:b9}
\frac{[\mbox{\bf S}_{D}^{f}]_{n0,00}}
{
\sum_{n'=1}^{\infty}
[\mbox{\bf S}_{D}^{f}]_{n'0,00}
}
=-\sum_{n'=1}^{\infty}M^{E*}_{n0,n'0}a_{n'0},
\end{equation}
and scaling Eq.~(\ref{eq:b3b}) is replaced by
\begin{equation}
\label{eq:b10}
c_{n0}(\omega)=
\frac{
\sum_{n'=1}^{\infty}
[\mbox{\bf S}_{D}^{f}]_{n'0,00}
}
{\nu^{(0)}}
c_{00}(\omega)a_{n0}.
\end{equation}
Now from Eq.~(\ref{eq:b8a}) we get
\begin{multline}
\label{eq:b11}
1=\left(
-i(\omega-\omega_{0})
-[\mbox{\bf S}_{D}^{f}]_{00,00}
+\frac{[\mbox{\bf K}]_{01,00}^{2}}
{\nu^{(0)}}a_{01}
\right.\\
\left.
-\frac{1}{\nu^{(0)}}
\sum_{n=1}^{\infty}
[\mbox{\bf S}_{D}^{f}]_{n0,00}
\sum_{n'=1}^{\infty}
[\mbox{\bf S}_{D}^{f}]_{n'0,00}a_{n'0}
\right)c_{00}(\omega),
\end{multline}
and, as a consequence, the effective Lorentzian width and shift:
\begin{multline}
\label{eq:b12}
\Gamma_{\rm eff}+i\Delta_{\rm eff}=
-[\mbox{\bf S}_{D}^{f}]_{00,00}
+\frac{[\mbox{\bf K}]_{01,00}^{2}}{\nu_{\vec{v}}}
\\
-\frac{1}{\nu^{(0)}}
\sum_{n=1}^{\infty}
[\mbox{\bf S}_{D}^{f}]_{n0,00}
\sum_{n'=1}^{\infty}
[\mbox{\bf S}_{D}^{f}]_{n'0,00}a_{n'0}.
\end{multline}

\bibliography{bib}

\end{document}